\documentclass[twocolumn]{aastex7}
\usepackage{color}
\usepackage{amsmath}
\usepackage{amssymb}
\usepackage{mathtools}
\usepackage{upgreek}
\usepackage{enumitem}
\usepackage{gensymb}
\usepackage{graphicx}
\usepackage{bm}
\usepackage{multirow}
\usepackage{mathrsfs}

\usepackage{tabularx}
\usepackage[T1]{fontenc}
\usepackage{cleveref}

\usepackage{xcolor}

\crefname{subsection}{subsection}{subsections}

\shortauthors{}
\shorttitle{}

\begin{document} 

\title{Revisiting discrete mappings for the orbital evolution of closely spaced planets}

\author[0009-0004-0844-173X]{Supawich Trongdee}
\affiliation{Department of Physics, Harvey Mudd College, Claremont, CA 91711, USA}
\email{jtrongdee@hmc.edu}



\author[0000-0002-9908-8705]{Daniel Tamayo}
\affiliation{Department of Physics, Harvey Mudd College, Claremont, CA 91711, USA}
\email{dtamayo@hmc.edu}

\begin{abstract}
At close orbital separations, the gravitational interaction between two planets becomes dominated by the close approaches each time the inner planet overtakes the outer.
This makes it possible to efficiently model the dynamics through a discrete mapping that evolves the system from each conjunction to the next.
We revisit the co-planar case of this map by \cite{Namouni96}, making minor simplifications to it based on more recent theoretical developments.
We then apply the mapping to investigate the role of higher order mean motion resonances in the onset of chaos at close separations, and find that our open-source \texttt{Python} implementation is over an order of magnitude faster than direct $N$-body integrations.

\end{abstract}

\section{Introduction} \label{sec:intro}


The chaotic dynamics of the planetary three-body problem with one star and two planets has a long and rich history.
Even specializing to co-planar orbits as we will do throughout this paper (and removing the center-of-mass degrees of freedom), one is still left with a formidable four degree-of-freedom system (2 for the $x$ and $y$ positions of each planet relative to the star). 
An expansive edifice of perturbative techniques has been developed over the past three hundred years to tackle this problem by averaging over fast oscillations and isolating dominant terms in the disturbing function expansion of the gravitational potential \citep[see, e.g.][]{Murray99, Morbidelli02}.

One regime in which the dynamics simplify significantly is the Hill limit of closely spaced orbits \citep{Hill1878}.
This is a particularly valuable setting in which to understand mean motion resonances (MMRs) between the planets' orbital periods, which are responsible for chaos and instabilities at close separations \citep{Wisdom80, Hadden18}.
At low eccentricities, the strongest resonances are first-order $p$:$p-1$ MMRs---a classic result that can be physically understood as due to higher order MMRs involving multiple conjunctions whose effects more nearly cancel out for progressively higher order resonances \citep{Jones26}.
Of these strong first-order MMRs, the most widely spaced is the 2:1 at a fractional semimajor axis separation $(a_2-a_1)/a_1 \approx 0.59$, with the 3:2, 4:3 etc. piling up at progressively closer spacings.
This regime appropriate to most strong MMRs simplifies the dynamics for at least two reasons.

First, in the closely spaced Hill limit, the gravitational interaction becomes dominated by the close approaches at conjunction where the interplanetary separation is minimized.
The gravitational force is so much larger during these close encounters that the interaction away from conjunction can be neglected, and the problem can be modeled as a discrete mapping from the effects of one conjunction to the next \citep{Duncan89} rather than a continuous problem. 

Second, in the Hill limit the dynamics of the two planets' center-of-mass and relative degrees of freedom become decoupled.
The center-of-mass degree of freedom is easily solved, and the non-trivial dynamics then only depends on the relative semimajor axis $a_2 - a_1$ and the relative eccentricity vector $\vec{e}_2 - \vec{e}_1$, where each eccentricity vector points toward its orbit's pericenter with a magnitude given by its orbital eccentricity  \citep{Henon86}.

The latter fact makes it possible to translate the problem of two closely spaced, massive planets on eccentric orbits to the simpler planar, circular restricted 3-body problem (PC3BP), in the following way \citep{Hadden18, Hadden19}.
First, the two bodies are kept at their original semimajor axes.
The planetary mass $m_p=m_1+m_2$ is then moved into the inner object (henceforth the ``planet" with subscript $p$), leaving the outer body as a test particle (no subscripts).
Finally, the planet is assigned a circular orbit $e_p=0$, while the test particle is initialized with the relative eccentricity vector $\vec{e} = \vec{e}_2 - \vec{e}_1$.
The unperturbed planetary orbit then never changes (analogous to the center-of-mass degree of freedom in the general problem), and the semimajor axis and eccentricity evolution in this PC3BP will match that of the relative semimajor axis and eccentricity in the more general problem of two massive planets on two elliptical orbits.
For a physical explanation of this correspondence, see \cite{Tamayo25}.

This makes the PC3BP the fundamental setting in which to understand MMRs, and throughout this paper we will specialize to an inner massive planet on a circular orbit with semimajor axis $a_p$, and an outer test particle at a semimajor axis $a$ with eccentricity $e$ (knowing one can easily translate quantities to the general case).

\cite{Duncan89} were the first to consider the orbital changes across a close encounter \citep{Henon86} and derive a discrete mapping from each conjunction to the next to model the dynamics of closely spaced planets.
Their mapping, valid to leading order in the eccentricity, was later generalized to higher order by \cite{Namouni96}.
This refinement turns out to be necessary for modeling higher order $p$:$p-q$ MMRs with $q\geq 2$ since otherwise the leading order effects cancel out perfectly at conjunctions for higher order MMRs \citep{Jones26}.

In this paper we revisit the high-order ``Namouni mapping" \citep{Namouni96}, introducing some minor simplifications in light of more recent theoretical developments.
Most importantly, the fact that the orbits are closely spaced implies that the ``crossing eccentricity" $e_c$ is small (i.e., the value at which the outer orbit's pericenter would touch the inner orbit)
\begin{equation}
    e_c = \frac{\lvert a-a_p\rvert}{a_p}\ll 1.
\end{equation}
\cite{Hadden18} and \cite{Hadden19} therefore argue that, rather than the absolute eccentricity, the important quantity is the normalized eccentricity
\begin{equation}
\tilde{e} \equiv \frac{e}{e_c}, \label{eq:etilde}
\end{equation}
i.e., the eccentricity as a fraction of the orbit-crossing value.
Expressing expansions in terms of $\tilde{e}$ leads to several simplifications in the Hill limit.
For example, the strengths and widths of MMRs of a given order collapse to a single universal expression \citep{Tamayo25}, rather than involving widely varying Fourier coefficients for each particular MMR in the traditional approach \citep[e.g.,][]{Murray99}.

The bulk of this paper is in the Appendix, where we work through the Namouni mapping. 
It may be valuable on a first reading to follow first the main text, where we begin in Sec.\:\ref{sec:mapping} by describing it and testing it against $N$-body integrations.
In Sec.\:\ref{sec:chaos} we then apply it to investigate the onset of chaos at close separations, exploiting the fact that we can use it to manually turn resonances on and off at higher orders.
We then conclude with a summary and some possible applications.
We provide an open-source implementation on Github.

\section{Conjunction mapping} \label{sec:mapping}

Approximating the planets' near-circular Keplerian orbits with the guiding center approximation \citep[see Sec.\:2.6 in][]{Murray99}, \cite{Namouni96} time-average the interaction potential between the two bodies across a single conjunction.
The orientation of the relative geometry between the two orbits is set by the longitude at which the two ellipses come closest together.
The effects of conjunctions therefore depend only on a single angle (henceforth the conjunction angle $\theta$) measuring how far the conjunction occurs relative to where the orbits are most closely separated (see Appendix \ref{b:Mapping} for details). 
Fourier expanding the effective conjunction potential $W$ in this angle\footnote{Like with the traditional disturbing function, the interaction potential's evenness in all the angles makes it possible to express $W$ only in terms of cosines} \citep{Namouni96},
\begin{equation}\label{namounipotentialmainpaper}
    W=\displaystyle\sum_{n=-\infty}^\infty W_n\cos(n\theta) \approx W_0+2\displaystyle\sum_{n=1}^M W_n\cos(n\theta),
\end{equation}
where in the second step we have used that $\cos(-n\theta) = \cos(n\theta)$ and exploited the fact that the potential is analytic and the Fourier coefficients decrease rapidly with $n$ to truncate the series at a particular order $M$. 

The Fourier coefficients $W_n = W_n(a_r, \tilde{e})$ depend on the relative semimajor axis and the normalized eccentricity (Eq.\:\ref{eq:etilde}). By appropriately writing $W$ in terms of action-angle variables, \cite{Namouni96} obtain the updates to each variable through partial derivatives with respect to its canonically conjugate pair.
This provides the updated longitude for the next conjunction, and the effects of subsequent encounters can be iteratively mapped to obtain the long-term dynamics. Appendix A provides the bulk of this paper, simplifying the mapping developed by \cite{Namouni96} by incorporating some of the insights from \cite{Hadden19} in the Hill limit, but can be omitted on a first reading that focuses on its applications.

For small eccentricities, the Fourier coefficients $W_n$ can also profitably be expanded as a power series in $\tilde{e}$ (Appendix A).
This reveals the well-known property that, to leading order in the eccentricity, the Fourier coefficient or `strength' of an $n$th order MMR scales as $\tilde{e}^n$ (i.e., a first-order MMR scales linearly with $\tilde{e}$, a second-order MMR as $\tilde{e}^2$ etc.).
For a physical explanation for the weakness of progressively higher order resonances in terms of the canceling effects of multiple conjunctions, see \cite{Jones26}.

\begin{figure}[htb!]
    \centering
    \includegraphics[width=\linewidth]{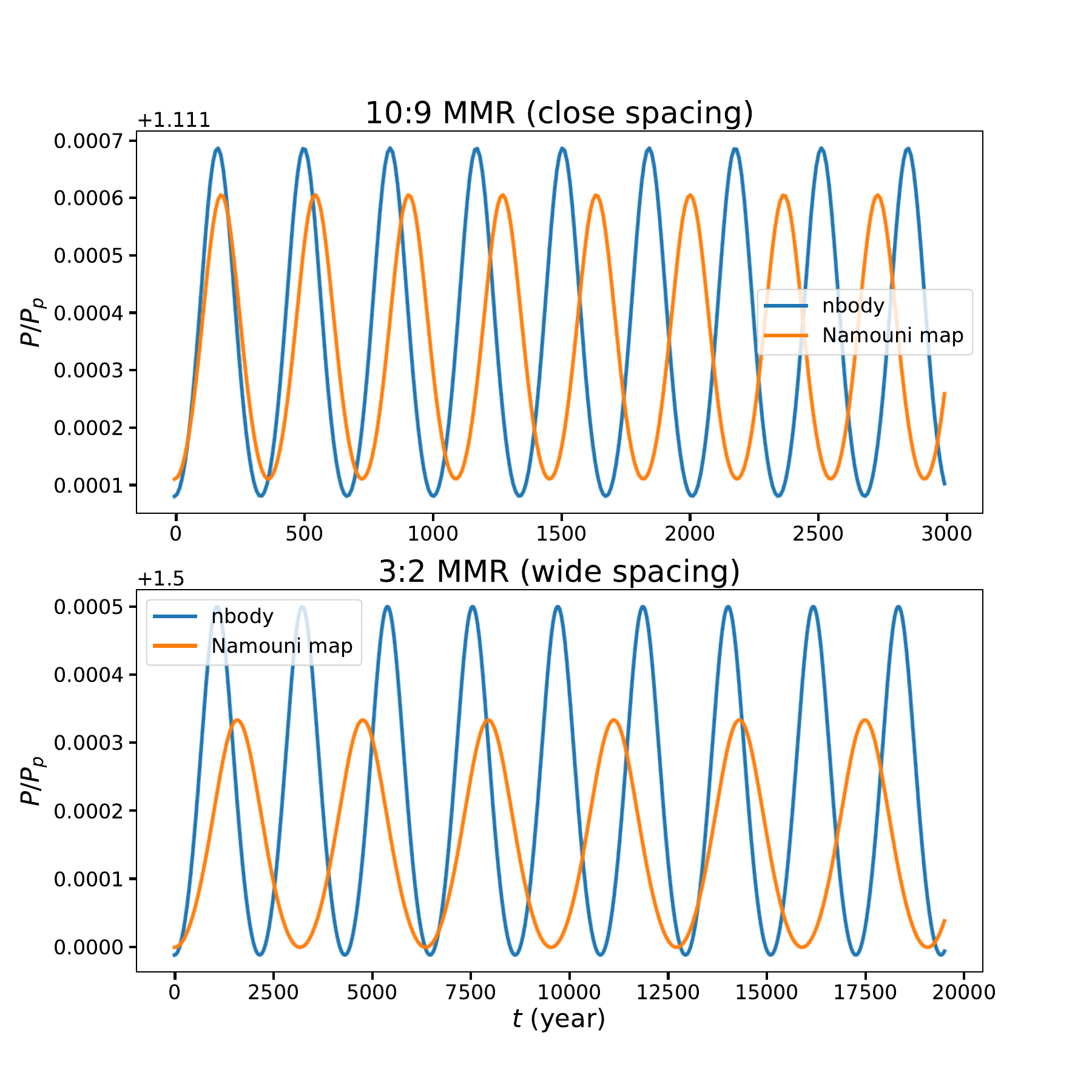}
    \caption{The evolution of the period ratio between the test particle and the massive planet in the Namouni mapping (orange) compared to $N$-body integrations using \texttt{REBOUND} (blue). The top plot shows the comparison when the orbits are more closely spaced with the initial period ratio of 10:9 over 300 conjunctions. The bottom plot shows the results over 6500 conjunctions when the initial spacing is 3:2.}
    \label{fig:avst_nbodyvsnamouni}
\end{figure}

\subsection{Tests}
To test our mapping implementation, we show in Fig.\:\ref{fig:avst_nbodyvsnamouni} the evolution of the period ratio using the Namouni mapping (orange) vs. $N$-body integrations using the \texttt{IAS15} integrator \citep{Rein15} in the \texttt{REBOUND} package \citep{Rein12}. 
The top panel is initialized in the Hill limit of close orbital spacings (a 10:9 period ratio), while the bottom panel shows a more widely separated setup where there begin to appear significant corrections (a 3:2 period ratio). In both cases we adopt a mass ratio of $\frac{m_p}{M_\star}=10^{-6}$ and a normalized eccentricity $\tilde{e}=0.005$ to emphasize that the mapping is accurate even for near-circular orbits. 

We see that in the top panel at close separations, the mapping matches the $N$-body integrations both in amplitude and frequency to within $\approx 10\%$. 
At the 3:2 in the bottom panel, corrections to the Hill limit result in discrepancies approaching $50\%$.

\begin{figure*}[htb!]
    \centering
    \includegraphics[width=\linewidth]{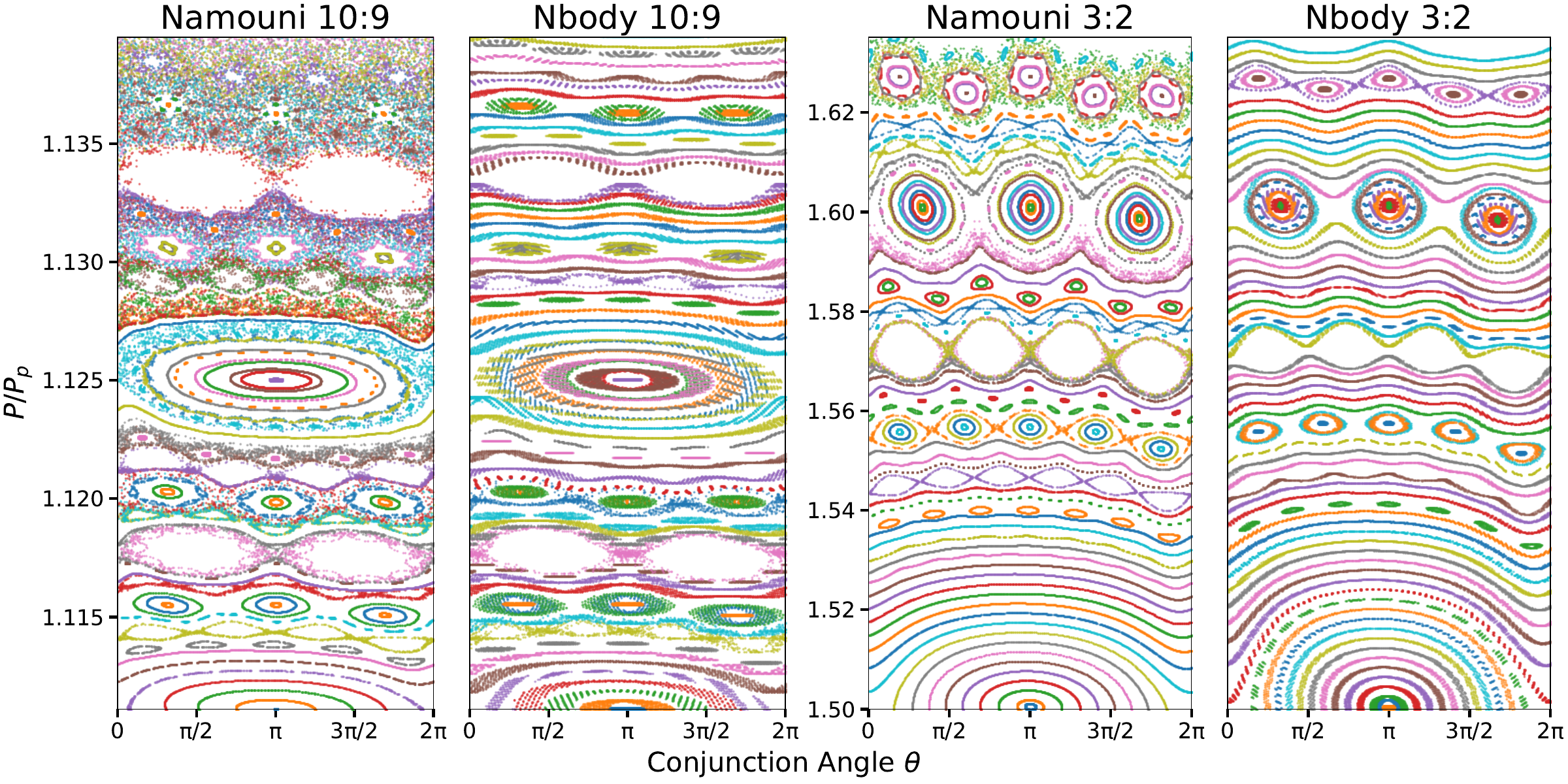}
    \caption{The surfaces of section in phase space of the conjunction angle and period ratio. The left two panels shows the phase space when the planets the mass ratio $\frac{m_p}{M_\star}=10^{-6}$. The leftmost panel is plotted using parameters from the Namouni map while the inner left panel uses parameters from REBOUND. The right two panels shows the phase space with the mass ratio $\frac{m_p}{M_\star}=10^{-4}$. The same normalized eccentricity $\tilde{e}=0.4$ is used in all panels.}
    \label{fig:sections_nbodyvsnamouni}
\end{figure*}

The dynamical evolution across different initial conditions is perhaps most effectively visualized as surfaces of section (Fig.\:\ref{fig:sections_nbodyvsnamouni}).
The left two panels consider a range of period ratios outward of the $10:9$ MMR with the Namouni mapping (leftmost panel) and $N$-body (second from left)\footnote{The $N$-body sections plot are taken between conjunctions, specifically at each time the mean longitudes of the two planets are separated by $\pi$ and the planets are approximately on opposite sides of the star.}
The right two panels show the corresponding surfaces of section outward from the $3:2$ MMR.
In order to obtain a mix of regular and chaotic trajectories in both the closely packed and more widely spaced configurations, we increase the initial normalized eccentricity for all initial conditions to $\tilde{e}=0.4$, and adopt a mass ratio of $\frac{m_p}{M_\star}=10^{-6}$ for the closely packed initial conditions near the $10:9$ MMR and a larger $\frac{m_p}{M_\star}=10^{-4}$ for the more widely spaced initial conditions near the $3:2$ MMR.

Various mean motion resonances are visible as islands at integer period ratios, where the number of islands matches the order of the resonance.
The edges (separatrices) of each of these resonant islands generically exhibits a chaotic layer that breaks up into points filling a finite 2-D space rather than following 1-dimensional curves like the regular trajectories do in the mapping panels (the $N$-body integrations have additional degrees of freedom that cause trajectories to blur out).

We can see that most features in Fig.\:\ref{fig:sections_nbodyvsnamouni} are consistent between the Namouni map and $N$-body integrations. 
The Namouni mapping requires one to choose an order to which to carry out the expansion in powers of the eccentricity, and following \cite{Namouni96} we choose an order $M=10$ for all our tests.

The runtime for the Namouni mapping (implemented in \texttt{Python}) is approximately 60 times faster than the $N$-body integration using \texttt{IAS15}, making it a valuable tool for investigating the long-term dynamics of closely separated planets. 

\begin{figure*}[htb!]
    \centering
    \includegraphics[width=\linewidth]{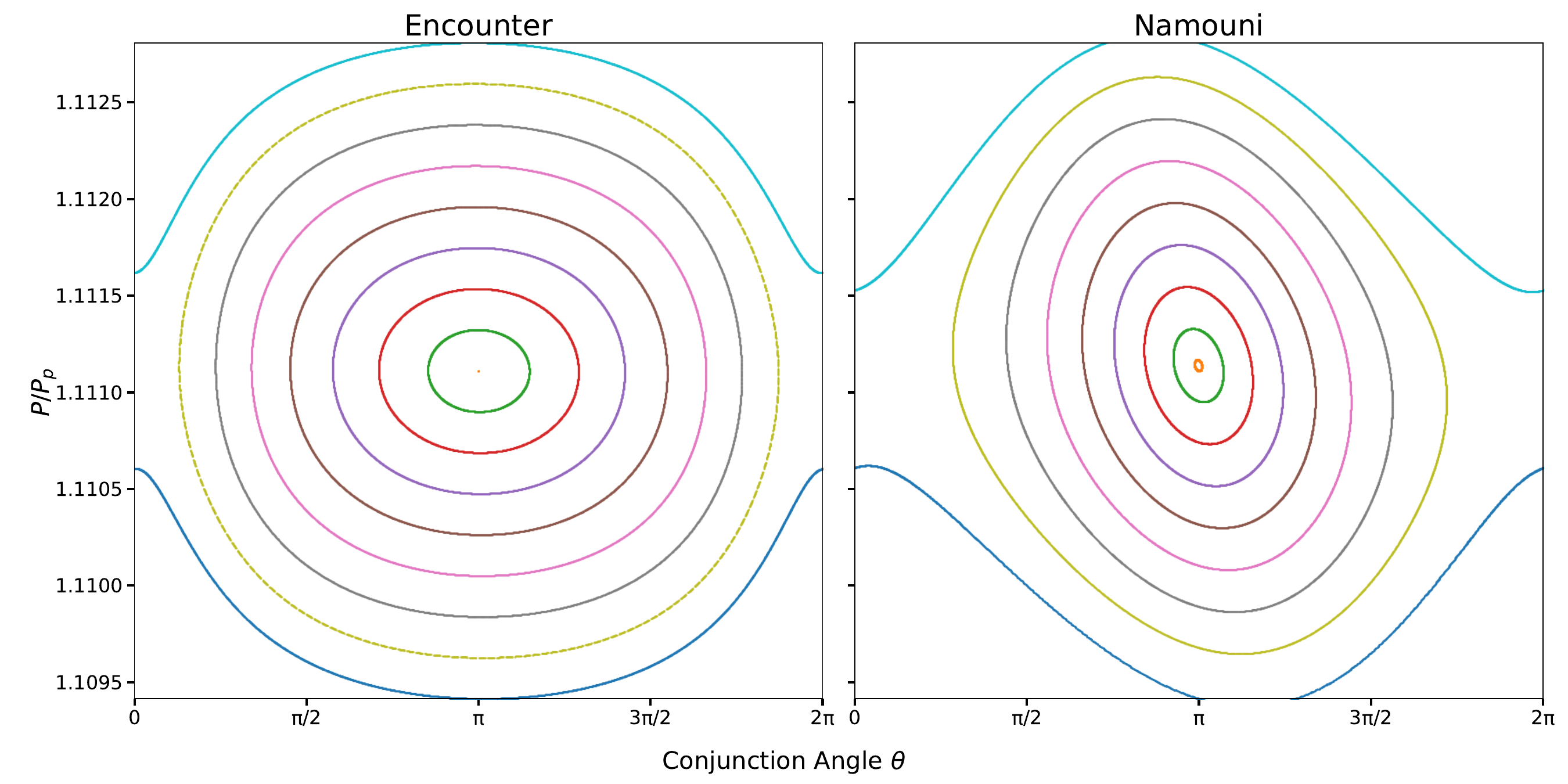}
    \caption{The surfaces of section in phase space of the conjunction angle and period ratio near the center of 10:9 resonance. The left panel is created by the EncounterMap class in celmech module, while the right panel is from the Namouni map.}
    \label{fig:sections_namounivsencounter}
\end{figure*}

Finally, the Namouni mapping can be compared with the symplectic \texttt{EncounterMap} implemented in \texttt{celmech} \citep{celmech}.
This mapping implements the disturbing function term for a single MMR from \cite{Hadden18} \citep[see also][]{Hadden19}. 
Specifically, it picks out an individual $n$ in Eq.\:\ref{namounipotentialmainpaper} corresponding to a single resonance, and evaluates its corresponding $W_n$ by numerically evaluating its integral representation rather than performing a series expansion in powers of the eccentricity \citep{Hadden18}. 
Figure \ref{fig:sections_namounivsencounter} plots a surface of section near the 10:9 resonance. 
We see that the main features, e.g., the width of the $10:9$ MMR matches to within 10\%.

\section{The Onset of Chaos} \label{sec:chaos}

\begin{figure*}[htb!]
    \centering
    \includegraphics[width=\linewidth]{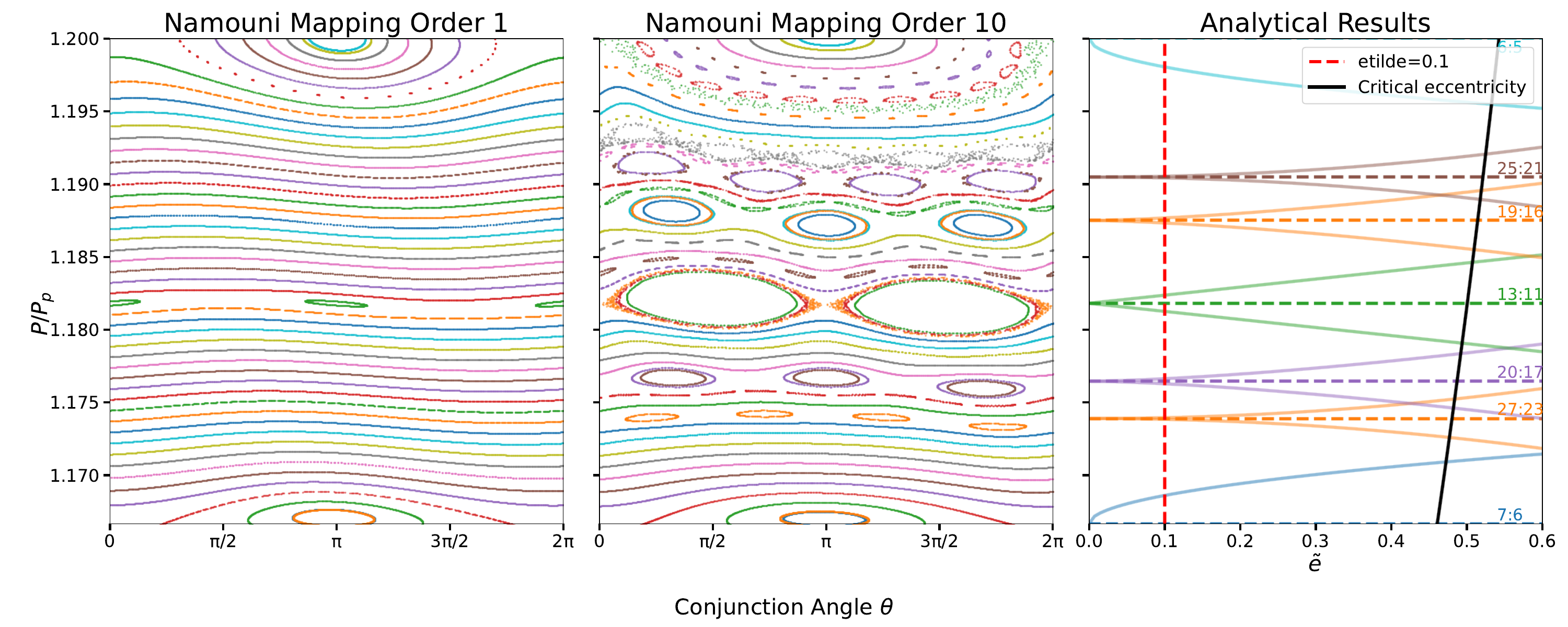}
    \caption{The surfaces of section in phase space of the period ratio and conjunction angle. The left panel shows the phase space from the first order Namouni map. The middle panel is from the Namouni map in the tenth order. The right panel plots the resonance widths as a function of normalized eccentricities. The same normalized eccentricity of 0.1 is used in both left and middle panels, indicated by the dashed red vertical line on the right panel. The slanted black line indicates the critical eccentricities for different period ratios.}
    \label{fig:lowecc_cateyes}
\end{figure*}

For given planetary masses, \cite{Wisdom80} derives a criterion for the onset of chaos of initially circular orbits at the critical semimajor axis separation that leads to the overlap of first-order MMRs (e.g., 5:4, 6:5, 7:6, etc.). 
This works well at low eccentricities, where only these first-order MMRs have significant widths and higher order MMRs ($p:p-q$ with $q>1$) are negligible.
\cite{Hadden18} generalizes this overlap criterion as a function of the relative eccentricity by incorporating higher order MMRs into the analysis, showing that they have comparable widths at large eccentricities and must be considered.
While obviously all resonances operate simultaneously in $N$-body integrations, the mapping of \cite{Namouni96} allows us to turn the effects of resonances at different orders on and off to study their behavior.

We consider pairs of planets with a total planetary mass of approximately two Earth masses ($m_p/M_\star=6\times10^{-6}$) with period ratios in the range between 7:6 ($P/P_p = 1.167$) and 6:5.
In the right panel of Fig.\:\ref{fig:lowecc_cateyes}, we plot the widths of the first through fourth order MMRs as a function of eccentricity using the analytical formula in the pendulum approximation from \cite{Tamayo25},
\begin{equation}
    \Delta_\text{max}=3A_q\sqrt{\frac{m_p}{M_\star}\tilde{e}^q}
\end{equation}
where $A_q$ is a numerical constant of order unity common to all $q$th order MMRs and $\Delta_\text{max}$ is the relative half-width of the resonance in terms of the period ratio, i.e., the separatrix separating resonant from non-resonant trajectories of the $p$:$p-q$ MMR lies at $\frac{P}{P_p} = \frac{p}{p-q} \Big(1 \pm \Delta_\text{max}\Big)$.

The black line in the right panel plots the approximate analytic expression for the chaotic boundary from \cite{Hadden18},
\begin{equation}\label{criticalecc}
    \tilde{e}_{\text{critical}}=\exp\Bigg[-2.2\Bigg(\frac{m_p}{M_\star}\Bigg)^{\frac{1}{3}}\Bigg(\frac{a_r}{a_p}\Bigg)^{-\frac{4}{3}}\Bigg],
\end{equation}
which we see closely matches the boundary at which adjacent resonances (on average) begin to overlap.

This right panel provides a simple view of the resonant landscape at a given mass. 
Additionally, specifying a normalized eccentricity corresponds to choosing a vertical red dashed line, whose intersections with each resonances specifies that MMR island's width at that eccentricity.
Specifying both the planetary masses and the relative eccentricity also allows us to generate surfaces of section like those shown above.
We consider a low-eccentricity (10\% of the orbit-crossing value) and a high-eccentricity (40\% of orbit-crossing) case in turn, which exhibit qualitatively different behavior.

\begin{figure*}[htb!]
    \centering
    \includegraphics[width=\linewidth]{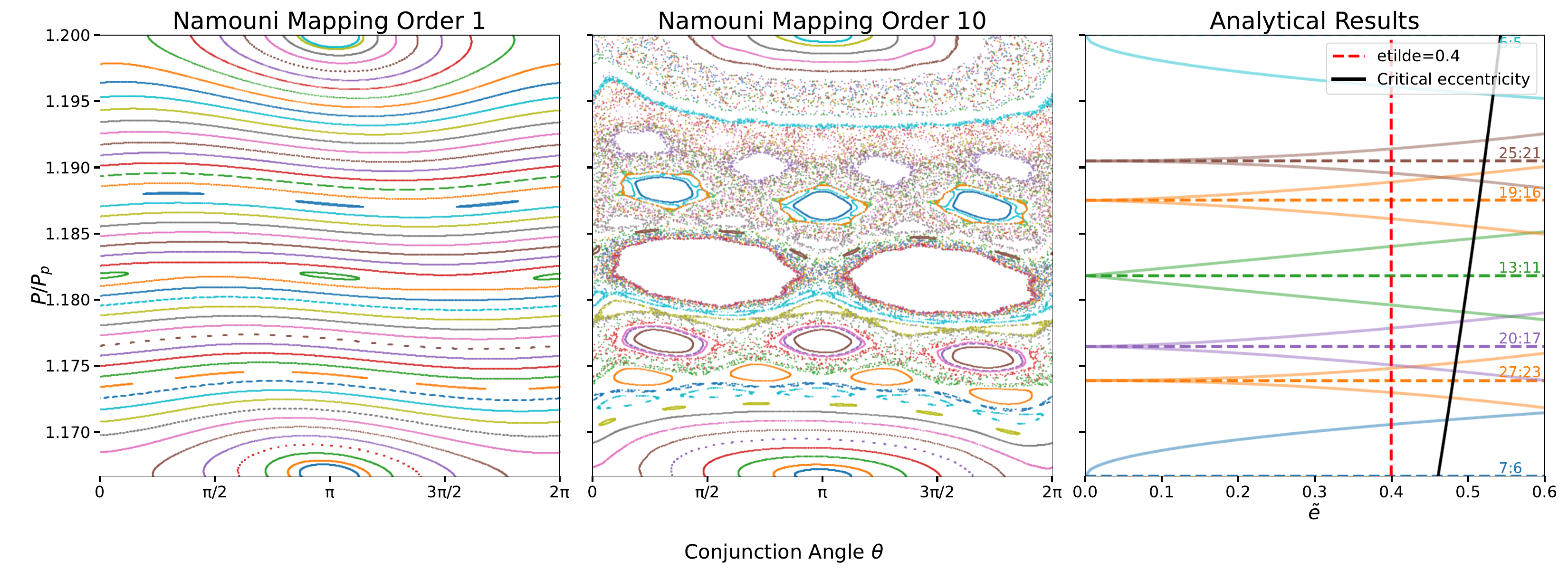}
    \caption{The surfaces of section in phase space of the period ratio and conjunction angle. The left panel shows the phase space from the first order Namouni mapping. The middle panel is from the Namouni mapping in the tenth order. The right panel plots the resonance widths as a function of normalized eccentricities. The same normalized eccentricity of 0.4 is used in both left and middle panels, indicated by the dashed red vertical line on the right panel. The slanted black line indicates the critical eccentricities for different period ratios.}
    \label{fig:hiecc_cateyes}
\end{figure*}

\subsection{Low Eccentricities}\label{lowecc}
We begin by exploring low-eccentricity orbits initialized at 10\% of the orbit-crossing eccentricity (dashed red line in the right panel of Fig.\:\ref{fig:lowecc_cateyes}).

The left and middle panels of Fig.\:\ref{fig:lowecc_cateyes} show surfaces of section initialized with this eccentricity (and planetary mass of two Earth masses), sharing the same $y$-axis as the right panel (i.e., varying the same initial period-ratio between the two planets).
The left panel shows a mapping only including first-order ($M=1$) MMRs, while the middle panel includes MMRs up to order $M=10$.
The widths of the resonant islands in the mapping in the left panel approximately match the analytic widths predicted by the intersection between the red line and the various resonances in the right panel. 
We see that while the $M=10$ mapping in the middle panel introduces a number of higher order resonance islands, they are separated widely enough from each other that the chaos is limited to very thin layers around each separatrix.
In this regime, the predictions from resonance overlap only considering first-order MMRs \citep{Wisdom80} will yield comparable predictions to ones including higher order MMRs \citep{Hadden18}.

\subsection{High Eccentricities}\label{highecc}

At high eccentricity, the resonances become wider, and higher order MMRs begin to overlap, leading to widespread chaotic regions (middle panel of Fig.\:\ref{fig:hiecc_cateyes}). 
By contrast, the resonances in the mapping including only first-order MMRs (left panel with $M=1$) remain well separated, and predict only regular trajectories.
This shows that at high eccentricities, where higher order resonances have significant widths (right panel of Fig.\:\ref{fig:hiecc_cateyes}), higher order MMRs play a central role in driving chaos through resonance overlap and must be considered \citep{Hadden18}.
Only for near-circular orbits where only first-order MMRs have significant widths\footnote{We note that this is not captured by the pendulum approximation used in plotting resonance widths in the right panel of Fig.\:\ref{fig:hiecc_cateyes}, and one must adopt the second fundamental model of resonance \citep{Henrard83}.} is the onset of chaos driven by the overlap of first-order MMRs \citep{Wisdom80}.


\section{Conclusion}
The discrete conjunction mapping developed by \cite{Duncan89} and refined by \cite{Namouni96} provides an efficient model with the minimum number of degrees of freedom required to reproduce the chaotic behavior in the three-body problem of two closely spaced planets orbiting a central star.
In this paper, we revisit this mapping and make minor simplifications to it appropriate to the Hill limit at close separations, and we use it to explore the onset of chaos in both the near-circular and eccentric limits.

Building on the classic result by \cite{Wisdom80} for the onset of chaos at the overlap of first-order MMRs for near-circular orbits, we reproduce the result from \cite{Hadden18} that at high eccentricities, the overlap of higher order resonances must additionally be considered.
In addition to reproducing the latter analytical work, the mapping additionally makes it possible to efficiently explore regular islands of stability inside MMRs among the surrounding chaotic sea that are difficult to calculate analytically.

The mapping is over an order of magnitude faster than direct $N$-body methods, and we hope can be a valuable tool for modeling the dynamics of closely spaced planets.
We release an open-source \texttt{Python} implementation of the mapping together with the scripts that generated each of the figures above at \url{https://github.com/Supa-wich/MappingPaper}.

\begin{acknowledgements}
We would like to thank Mickey Teekamongkol and Lizzy Jones for their early contributions to this work.
\end{acknowledgements}

\appendix
\section{The Mapping Equations for Close Encounters} \label{b:Mapping}

From \cite{Namouni96}, the effective potential $W$ can be expanded as a Fourier series:
\begin{equation}\label{namounipotential}
    W=W_0+2\displaystyle\sum_{n=1}^M W_n\cos(n\theta)
\end{equation}
of the finite order $M$. The coefficient $W_n$ is
\begin{equation}
    W_n = \frac{1}{a_r}\displaystyle\sum_{\lvert n\rvert \leq p\leq M}W_n^{p,0}(\frac{e_r}{a_r})^p
\end{equation}
where $p$ shares the same parity as $n$.

At the Hill limit, $a_r\approx e_c$, allowing us to rewrite the coefficients in terms of the normalized eccentricity:

\begin{equation}
    W_n = \frac{1}{a_r}\displaystyle\sum_{\lvert n\rvert \leq p\leq M}W_n^{p,0}\tilde{e}^p
\end{equation}

In the work of \cite{Namouni96}, $(I,\phi)$ and $(K,\lambda)$ are the action-angle variable pairs for the coplanar planets. The two actions are related given the conserved energy $H_0=I-K$. The relationships between these two pairs of action-angle variables and their physical counterparts are shown in Table \ref{tab:parameters}.
\begin{table}[]
    \centering
    \begin{tabular}{c|c}
        Variable & Relationship to Physical Parameter \\
        $I$ & $\frac{(e_r/\epsilon)^2}{2}$ \\
        $K$ & $\frac{3(a_r/\epsilon)^2}{8}$\\
        $\lambda$ & Conjunction longitude \\
        $\phi$ & Pericenter longitude
    \end{tabular}
    \caption{List of orbital parameters used in the mapping by \cite{Namouni96} in terms of action-angle variables.}
    \label{tab:parameters}
\end{table}

The conjunction longitude in Namouni's convention has a phase difference $\pi$ from the common pericenter longitude denoted by $\varpi$ in other findings such as \cite{Duncan89}. The conjunction angle in the Namouni mapping is then defined as $\theta_{\text{Namouni}}=\lambda-\phi$ instead of $\lambda-\varpi$. This means \cite{Namouni96} measure the conjunction angle when the planets are furthest apart. In our paper, we will make another $\pi$ shift so that it is measured at the conjunction, i.e. $\theta=\theta_{\text{Namouni}}-\pi$ which gives us the new coefficients due to the phase shift of $n\pi$ in the cosines

\begin{equation}
    W_n = \frac{(-1)^n}{a_r}\displaystyle\sum_{\lvert n\rvert \leq p\leq M}W_n^{p,0}\tilde{e}^p
\end{equation}

The potential becomes
\begin{equation}\label{newpotential}
    W=W_0+2\displaystyle\sum_{n=1}^M \frac{(-1)^n}{a_r}(\displaystyle\sum_{\lvert n\rvert \leq p\leq M}W_n^{p,0}\tilde{e}^p)\cos(n\theta)
\end{equation}

and we define $\theta=\lambda-\varpi$.

The updates to the action-angle variables at the $n+1$th conjunction (in Namouni's notation) are

\begin{align}
    I_{n+1}&=I_n+\partial_\phi W(\phi_n,\lambda_n,I_{n+1},K_{n+1}) \\
    \phi_{n+1}&=\phi_n-\partial_I W(\phi_n,\lambda_n,I_{n+1},K_{n+1})  \\
    K_{n+1}&=K_n-I_n+I_{n+1} \\
    \lambda_{n+1}&=\lambda_n+\partial_K W(\phi_n,\lambda_n,I_{n+1},K_{n+1})+\frac{4\pi}{3\epsilon\sqrt{8K_{n+1}/3}}
\end{align}

The third term of the update for $\lambda_{n+1}$ is the angle at which the planet sweeps after one synodic period, which must be considered since both planets are orbiting the star and interacting with each other at the same time. The updates are symplectic due to having the actions at the $n+1$th conjunction on both sides of the mapping equation. This mapping can be implemented in the code by iterating the above updates to some accepted tolerance like any implicit numeral integration scheme.

With the perturbing potential rewritten as a function of $\tilde{e}$, we can define the new action $\tilde{I}=\frac{\tilde{e}^2}{2}$ that is still conjugated with $\phi$ and is just a rescaling of $I$. That is, $I = \tilde{I}(\frac{a_r}{\epsilon})^2$. In our notation, this is conjugated with $\varpi$, so we can apply the chain rule for the updates:
\begin{equation}
    \frac{\partial W}{\partial\phi}= \frac{\partial W}{\partial\varpi} \frac{\partial \varpi}{\partial\phi}=\frac{\partial W}{\partial\varpi} \frac{\partial (\phi+\pi)}{\partial\phi}=\frac{\partial W}{\partial\varpi}
\end{equation}
and 
\begin{equation}
    \frac{\partial W}{\partial I}= \frac{\partial W}{\partial\tilde{I}} \frac{\partial \tilde{I}}{\partial I}=\frac{\partial W}{\partial\tilde{I}} \frac{\partial (\frac{\epsilon}{a_r})^2I}{\partial I}=(\frac{\epsilon}{a_r^2})\frac{\partial W}{\partial\tilde{I}}
\end{equation}
so that 
\begin{align}
    \tilde{I}_{n+1}&=\frac{\epsilon}{a_r^2}(I_n+\partial_\phi W(\phi_n,\lambda_n,I_{n+1},K_{n+1})) \\
    &=\tilde{I}_n+\partial_\varpi W(\varpi_n,\lambda_n,\tilde{I}_{n+1},K_{n+1})\nonumber
\end{align}
and
\begin{equation}
    \varpi_{n+1}=\varpi_n-(\frac{\epsilon}{a_r^2})\partial_{\tilde{I} }W(\varpi_n,\lambda_n,\tilde{I}_{n+1},K_{n+1})
\end{equation}

We can expand $W$ at $M=4$ as follows
\begin{align}\label{fourthorder}
    W&=\frac{1}{a_r}\displaystyle((-1)^0W_0^{0,0}+2(-1)^1W_1^{1,0}\sqrt{2\tilde{I}}\cos\theta+2(-1)^2(W_0^{2,0}+2W_2^{2,0}\cos 2\theta)\tilde{I}\\
    &\quad +2(-1)^3(W_1^{3,0}\cos\theta+W_3^{3,0}\cos 3\theta)(2\tilde{I})^{3/2}+4(-1)^4(W_0^{4,0}+2W_2^{4,0}\cos 2\theta +2W_4^{4,0}\cos 4\theta)\tilde{I}^2\displaystyle) \nonumber
\end{align}
which gives us the update to the action-angle variables for eccentricity:
\begin{align}
    \tilde{I}_{n+1}&=\tilde{I}_n+\frac{2W_1^{1,0}}{a_r}\sqrt{2\tilde{I}_{n+1}}\sin\theta_n-(4W_2^{2,0}\sin2\theta_n)\frac{2\tilde{I}_{n+1}}{a_r}\label{Iupdate}\\
    &\quad+(2W_1^{3,0}\sin\theta_n+6W_3^{3,0}\sin3\theta_n)\frac{(2\tilde{I}_{n+1})^{3/2}}{a_r}-(4W_2^{4,0}\sin2\theta_n+8W_4^{4,0}\sin4\theta_n)\frac{4\tilde{I}_{n+1}^2}{a_r}\nonumber \\
    \varpi_{n+1}&=\varpi_n+(\frac{\epsilon}{a_r^2})(\frac{\sqrt{2}W_1^{1,0}}{a_r^3}\cos\theta_n\frac{1}{\sqrt{\tilde{I}_{n+1}}}-(W_0^{2,0}+2W_2^{2,0}\cos2\theta_n)\frac{2}{a_r^3} \label{phiupdate}\\
    &\quad+3\cdot2^{3/2}(W_1^{3,0}\cos\theta_n+W_3^{3,0}\cos3\theta_n)\frac{\tilde{I}^{1/2}}{a_r^3}-8(W_0^{4,0}+2W_2^{4,0}\cos2\theta_n+2W_4^{4,0}\cos4\theta)\frac{\tilde{I}}{a_r^3})\nonumber
\end{align}
In the low-eccentricity limit, the term $\frac{1}{\sqrt{\tilde{I}}}$ in $\varpi$ causes the mapping to blow up. One solution to this is to introduce the correction to the first order term by the singular potential (in terms of our new action $\tilde{I}$)
\begin{equation}
    W_{\text{sing}}=\frac{2W_1^{1,0}}{a_r^3}\sqrt{2\tilde{I}}\cos\theta=-\frac{2W_1^{1,0}}{a_r^3}(\tilde{h}\cos\lambda+\tilde{k}\sin\lambda)
\end{equation}
where $\tilde{h}=\sqrt{2\tilde{I}}\cos\varpi$ and $\tilde{k}=\sqrt{2\tilde{I}}\sin\varpi$. The singular potential can in fact be expanded further as suggested in Section 3.3 of \cite{Namouni96}, but we need only the first order since it is the only problematic term at low eccentricity.
Using this, we update the eccentricity components $(\tilde{h},\tilde{k})$ instead of the original first order update on $(\tilde{I},\varpi)$:
\begin{align}  \tilde{h}'&=\tilde{h}_n+\partial_{\tilde{k}}W_{\text{sing}}(\tilde{h}_{n+1},\tilde{k}_n,\lambda_n) \\
    &=\tilde{h}_n-\frac{2W_1^{1,0}}{a_r^3}\sin\lambda_n \nonumber\\
    \tilde{k}'&= \tilde{k}_n-\partial_{\tilde{h}}W_{\text{sing}}( \tilde{h}_{n+1}, \tilde{k}_n,\lambda_n) \\
    &= \tilde{k}_n+\frac{2W_1^{1,0}}{a_r^3}\cos\lambda_n \nonumber
\end{align}
Again, our ``half-step'' update here has the opposite signs compared to Namouni's work due to the $\pi$ phase shift to our defined conjunction angle.
We can convert back to $(\tilde{I},\varpi)$ coordinates by
\begin{align}
    \tilde{I}'&=\frac{\tilde{h}'^2+\tilde{k}'^2}{2} \\
    \varpi'&=\arctan(\frac{\tilde{k}'}{\tilde{h}'})
\end{align}
which are the updates in place of the first order term. The intermediate update in the conjunction angle is $\theta'=\lambda_n-\varpi'$.
Now we are ready to add higher order updates: 
\begin{align}
    \tilde{I}_{n+1}&=\tilde{I}'-(4W_2^{2,0}\sin2\theta')\frac{2\tilde{I}_{n+1}}{a_r}\\
    &\quad +(2W_1^{3,0}\sin\theta'+6W_3^{3,0}\sin3\theta')\frac{(2\tilde{I}_{n+1})^{3/2}}{a_r}-(4W_2^{4,0}\sin2\theta'+8W_4^{4,0}\sin4\theta')\frac{4\tilde{I}_{n+1}^2}{a_r}\nonumber \\
    \varpi_{n+1}&=\varpi'-(\frac{\epsilon}{a_r^2})((W_0^{2,0}+2W_2^{2,0}\cos2\theta')\frac{2}{a_r^3} \\
    &\quad-3\cdot2^{3/2}(W_1^{3,0}\cos\theta'+W_3^{3,0}\cos3\theta')\frac{\tilde{I}_{n+1}^{1/2}}{a_r^3}+8(W_0^{4,0}+2W_2^{4,0}\cos2\theta'+2W_4^{4,0}\cos4\theta')\frac{\tilde{I}_{n+1}}{a_r^3})\nonumber
\end{align}
In summary, the discrete update of the orbital parameters includes the following steps
\begin{align}
    \tilde{h}_n&=\sqrt{2\tilde{I}_n}\cos\varpi_n\\
    \tilde{k}_n&=\sqrt{2\tilde{I}_n}\sin\varpi_n\\
    \tilde{h}'&=\tilde{h}_n-\frac{2W_1^{1,0}}{a_r^3}\sin\lambda_n\\
    \tilde{k}'&=\tilde{k}_n+\frac{2W_1^{1,0}}{a_r^3}\cos\lambda_n \\
    \tilde{I}'&=\frac{\tilde{h}^2+\tilde{k}^2}{2} \\
    \varpi'&=\arctan(\frac{k'}{h'})\\
    \theta'&=\lambda_n-\varpi'\\
    \tilde{I}_{n+1}&=\tilde{I}'-(4W_2^{2,0}\sin2\theta')\frac{2\tilde{I}_{n+1}}{a_r}\\
    &\quad +(2W_1^{3,0}\sin\theta'+6W_3^{3,0}\sin3\theta')\frac{(2\tilde{I}_{n+1})^{3/2}}{a_r}-(4W_2^{4,0}\sin2\theta'+8W_4^{4,0}\sin4\theta')\frac{4\tilde{I}_{n+1}^2}{a_r}\nonumber \\
    \varpi_{n+1}&=\varpi'-(\frac{\epsilon}{a_r^2})((W_0^{2,0}+2W_2^{2,0}\cos2\theta')\frac{2}{a_r^3} \\
    &\quad-3\cdot2^{3/2}(W_1^{3,0}\cos\theta'+W_3^{3,0}\cos3\theta')\frac{\tilde{I}_{n+1}^{1/2}}{a_r^3}+8(W_0^{4,0}+2W_2^{4,0}\cos2\theta'+2W_4^{4,0}\cos4\theta')\frac{\tilde{I}_{n+1}}{a_r^3})\nonumber \\
    K_{n+1}&=(\tilde{I}_{n+1}-\tilde{I}_n)(a_r/\epsilon)^2+K_n \\
    \lambda_{n+1}&=\lambda_n+\frac{4\pi}{3\epsilon\sqrt{8K_{n+1}/3}}\label{lambdaupdate}
\end{align}

One can analyze the dynamics of the system by considering the phase space of $a_r/\epsilon$ and $\theta$ as demonstrated in \cite{Namouni96}. However, to more easily visualize MMRs (which fall at integer period ratios) in our surfaces of section like those in our main paper, we will switch to plotting the period ratio on the y-axis,
\begin{equation}
\frac{P}{P_p} = 1 + \frac{P-P_p}{P_p} \approx 1 + \frac{3}{2} \frac{a - a_p}{a_p} = 1 + \frac{3}{2}a_r. 
\end{equation}

\section{Exact Conjunction Longitude Update}\label{c:exactlambda}
\begin{figure*}[htb!]
    \centering
    \includegraphics[width=\linewidth]{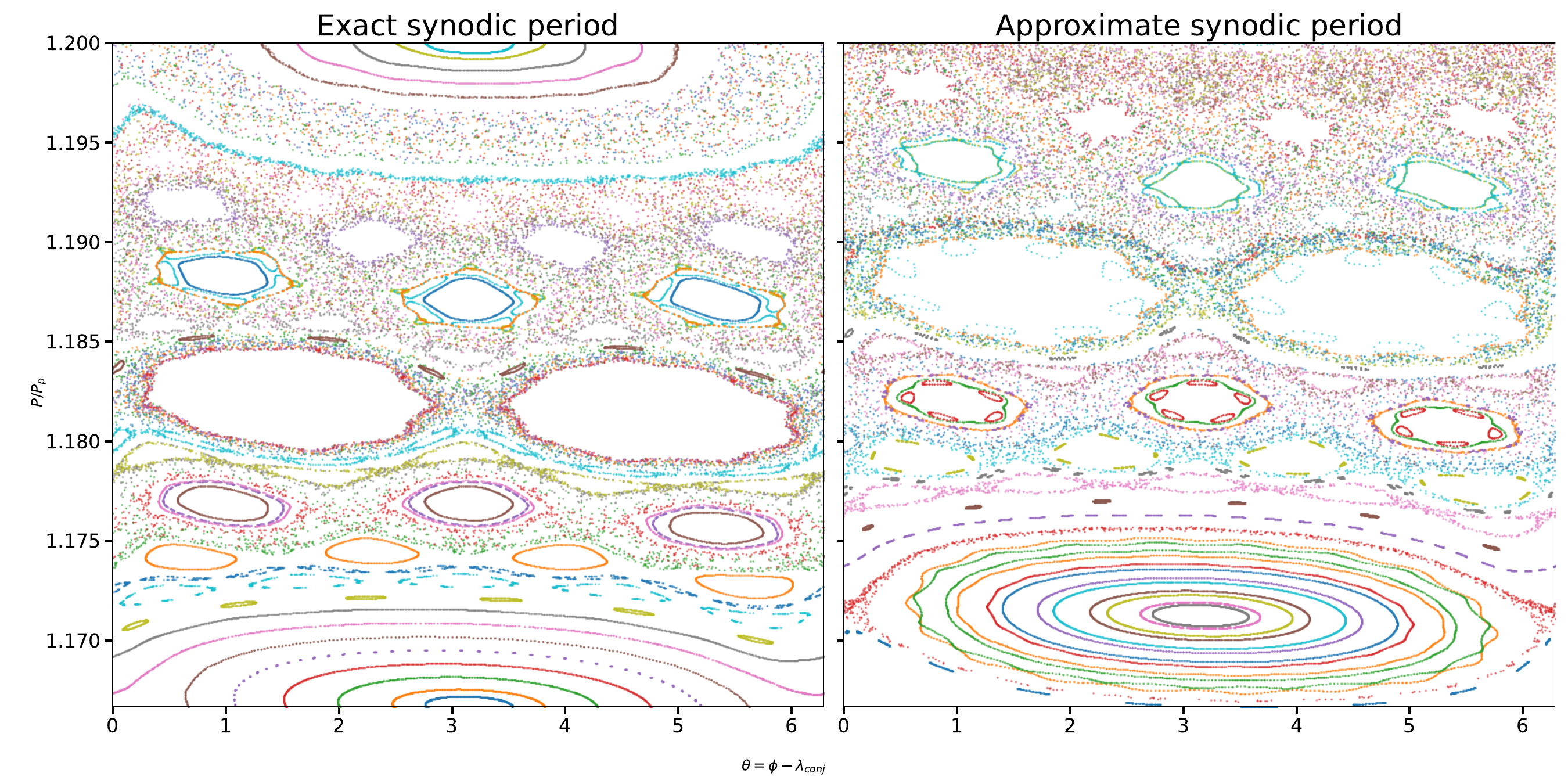}
    \caption{The surfaces of section in phase space of the period ratio and conjunction angle. Both shows the phase space from the tenth order Namouni mapping with the exact same initial conditions from Fig.\:\ref{fig:hiecc_cateyes}. The left panel uses the exact synodic period calculation from \cite{Duncan89}, while the right panel uses the approximated version as derived in \cite{Namouni96}.}
    \label{fig:exactvsapprox_cateyes}
\end{figure*}

Recall Eq. \ref{lambdaupdate} for the mapping update in the longitude at the $n+1$th conjunction:
\begin{equation*}
    \lambda_{n+1}=\lambda_n+\frac{4\pi}{3\epsilon\sqrt{8K_{n+1}/3}}
\end{equation*}
However, the synodic period (second term) in the update is the Hill limit approximation. Using this will not change the dynamics of the systems in phase space but can cause the deviation of the period ratio as seen in Fig.\:\ref{fig:exactvsapprox_cateyes} due to the cumulative difference between the approximate version of the conjunction angle and the actual value from the simulation. In the left panel, the center of the 7:6 resonance stays exactly at 7:6, but if the approximation is used, the center of stable islands around 7:6 is shifted up to be about 1.172.

To address that, we can use the exact synodic period without the Hill limit approximation from \cite{Duncan89} in their version of the mapping. Their version of the update in longitude reads
\begin{equation}
    \lambda_{n+1}=\lambda_n+2\pi f(\varepsilon_{n+1})
\end{equation}
where the curly epsilon $\varepsilon=a_r$ in our paper and the function $f$ can be rewritten in our notation as
\begin{equation}\label{f_ar}
    f(a_r)=\lvert(1+a_r)^{-3/2}-1\rvert^{-1}
\end{equation}
Now we will show that $2\pi f(a_r)$ is approximately $\frac{4\pi}{3\epsilon\sqrt{8K/3}}$, implying that the longitude updates by \cite{Duncan89} and \cite{Namouni96} are equivalent at the Hill limit. Since, $a_r\ll 1$ by our Hill limit assumption, we can simply Taylor expand Eq. \ref{f_ar}:
\begin{equation}
    f(a_r)\approx\lvert(1-\frac{3}{2}a_r)-1\rvert^{-1}=\frac{2}{3a_r}
\end{equation}
Then we rewrite this in terms of $K$ by its definition:
\begin{align}
    f(a_r)&\approx\frac{2}{3a_r} \\
    &= \frac{2}{3\epsilon\sqrt{8K/3}}
\end{align}
So,
\begin{align}
    \lambda_{n+1}&=\lambda_n+2\pi f(\varepsilon_{n+1}) \\
    &\approx\lambda_n+2\pi \frac{2}{3a_{r,n+1}} \\
    &= \lambda_n+\frac{4\pi}{3\epsilon\sqrt{8K_{n+1}/3}}
\end{align}


\bibliographystyle{aasjournal}
\bibliography{main}     
\end{document}